\documentclass[qsecnum,amsmath,preprintnumbers,superscriptaddress,nofootinbib,aps,prd,11pt,a4paper]{revtex4-1}
\pdfoutput=1
\usepackage[utf8]{inputenc}
\usepackage[english]{babel}
\usepackage{amsmath}
\usepackage{amsfonts}
\usepackage{amssymb}
\usepackage{amsthm}
\usepackage{hyperref}
\usepackage{color}
\usepackage{tikz}
\usepackage{feynmp-auto}
\usepackage{bm}
\usepackage{graphicx}
\usetikzlibrary{matrix}

 \def\be{\begin{equation}}
\def\ee{\end{equation}}
 \def\ba{\begin{align}}
\def\ea{\end{align}}
\def\bea{\begin{eqnarray}}
\def\eea{\end{eqnarray}}

\def\m{\mu}

\def\Z{\frac{(1+6\xi_h)\xi_\chi}{(1+6\xi_\chi)\xi_h}}
\def\E{\frac{\xi_\chi}{\xi_h}}

\begin{document}
\preprint{INR-TH-2019-011}
\title{{\bf To Positivity and Beyond,\\ where Higgs-Dilaton Inflation has never gone before}}
\author{Mario Herrero-Valea}
\email[]{mario.herrerovalea@epfl.ch}
\address{Institute of Physics, LPPC. \'Ecole Polytechnique F\'ed\'erale de Lausanne\\ CH-1015 Lausanne, Switzerland}

\author{Inar Timiryasov}
\email[]{inar.timiryasov@epfl.ch}
\address{Institute of Physics, LPPC. \'Ecole Polytechnique F\'ed\'erale de Lausanne\\ CH-1015 Lausanne, Switzerland}

\author{Anna Tokareva}
\email[]{tokareva@ms2.inr.ac.ru}
\address{Institute for Nuclear Research of Russian Academy of Sciences\\
117312 Moscow, Russia}

\begin{abstract}
We study the consequences of (beyond) positivity of scattering amplitudes in the effective field theory description of the Higgs-Dilaton inflationary model. By requiring the EFT to be compatible with a unitary, causal, local and Lorentz invariant UV completion, we derive constraints on the Wilson coefficients of the first higher order derivative operators. We show that the values allowed by the constraints are consistent with the phenomenological applications of the Higgs-Dilaton model.
\end{abstract}

\maketitle
\newpage
\tableofcontents
\newpage
\section{Introduction}
Effective field theories (EFTs) are ubiquitous in theoretical physics. They represent the success of a very profound ordering principle of Nature -- the fact that low energy physics can be described solely in terms of light degrees of freedom, encoding the effect of heavy particles and unknown ultraviolet (UV) interactions in effective operators of increasing dimension, suppressed by a cut-off $\Lambda$. Retaining a finite number of terms of this otherwise infinite tower of operators, we can describe phenomena at energies below the cut-off with an increasing accuracy, without the knowledge of a full UV description.

However, EFTs are generically unable to capture non-analytical features of their UV completion \cite{Cohen:2019wxr}, and it could be the case that a perfectly healthy derivative expansion of a Lagrangian could never correspond to the low-energy expansion of an also healthy UV complete theory. In other words, not all the possible EFTs that we can write will correspond to the low-energy expansion of a UV complete theory satisfying some desirable properties -- locality, unitarity, causality and Lorentz invariance \cite{GellMann:1954db,Goldberger:1955zz,Eden:1966dnq,Martin:944356}. These principles are deeply engraved in our understanding of Nature and we have strong compelling reasons to expect any reasonable UV completion to contain them as basic ingredients. In a Quantum Field Theory (QFT), they are encoded in properties of the S-matrix such as crossing symmetry and polynomial boundedness of scattering amplitudes and thus they can be tested at a purely mathematical level provided that we can compute the corresponding contributing amplitudes. The most typical example being the 2-to-2 scattering amplitude in the forward limit. In particular, with the knowledge of this element, unitarity can be used in the form of the optical theorem and together with Lorentz invariance of the EFT, to derive a connection between infrared (IR) and UV physics in the form of \emph{positivity bounds} \cite{Adams:2006sv}. Some combinations of Wilson coefficients in the EFT are constrained to be positive. Coefficients of the wrong sign can never be derived from a unitary UV completion. 

More recently it has been shown that by appealing to the explicitly positive character of the scattering cross-section for 2-to-2 processes and to the aforementioned approximation of the UV completion by the EFT within a certain range of energies, one can go \emph{beyond positivity} \cite{Bellazzini:2017fep}. That is, we can constrain the Wilson coefficients not just to be positive, but to be larger than a second quantity which can be also computed in the EFT without any knowledge of the UV completion. Thus, one can establish a strong bound on the consistency of the EFT itself without any extra external or unknown ingredients. Examples of the use of positivity and beyond positivity bounds, as well as efforts to generalize them, are multiple. An incomplete list includes the proof of the $a$-theorem \cite{Komargodski:2011vj,Luty:2012ww}, the study of chiral perturbation theory \cite{Manohar:2008tc}, composite Higgs \cite{Low:2009di,Urbano:2013aoa}, different models of modified and quantum gravity \cite{Bellazzini:2015cra,Cheung:2016yqr, Bonifacio:2016wcb,Keltner:2015xda,deRham:2017imi}, higher-spin theories \cite{Bellazzini:2019bzh}, the null energy condition \cite{Nicolis:2009qm}, non-local theories \cite{Tokuda:2019nqb}, the Weak Gravity Conjecture \cite{Cheung:2014ega,Cheung:2014vva,Bellazzini:2019xts}, conformal field theory \cite{Hartman:2015lfa,Alday:2016htq}, estimation of cosmological parameters \cite{Melville:2019wyy}, and the EFT of inflation \cite{Baumann:2015nta,Croon:2015fza}. Further applications outside the forward limit and for massless particles of integer spin, together with several interesting references, can be found in \cite{Bellazzini:2019xts,Bellazzini:2016xrt,deRham:2018qqo,deRham:2017zjm,deRham:2017avq}. In a broad sense, this direction of research parallels the recent Swampland program in String Theory \cite{Palti:2019pca} with a similar philosophy -- not all EFT's can be derived from a reasonable UV completion (String Theory in their case).

A field where the formulation of EFTs is particularly fruitful is the early Universe Cosmology. In particular, due to the meager data at our disposal about the inflationary epoch, we cannot formulate a precise theory of inflation but instead there exist several successful models which are able to reproduce our main proxy of information -- the spectrum of scalar perturbations of the Cosmic Microwave Background \cite{Akrami:2018odb,Martin:2013tda}. Moreover, since any inflationary model needs to include gravity, they are inevitably EFTs due to the non-renormalizable character of gravitational interactions \cite{Donoghue:1994dn,Donoghue:2017pgk}.

A particular successful model of inflation is Higgs inflation \cite{Bezrukov:2007ep,Bezrukov:2010jz,Rubio:2018ogq}, where the Standard Model Higgs boson is coupled non-minimally to gravity and takes the role of the inflaton during the early epochs of the Universe. If the value of the non-minimal coupling $\xi_h$ is sufficiently large, this model is able to provide a successful inflationary epoch and a graceful exit\footnote{The gracefulness of this exit has been however recently questioned. See \cite{DeCross:2016cbs,Ema:2016dny,Gorbunov:2018llf,Bezrukov:2019ylq,He:2018mgb} and references therein.} into the standard hot Big Bang model \cite{Bezrukov:2011sz}. This is usually described in the so-called Einstein frame of the theory, where the non-minimal coupling is removed by a field redefinition in the form of a conformal rescaling.

Higgs Inflation (HI) can be embedded into a larger scenario known as Higgs-Dilaton Inflation (HDI) \cite{Bezrukov:2012hx,Casas:2017wjh,GarciaBellido:2011de}, where we add an extra scalar field $\sigma$ (the dilaton) whose non-minimal coupling to gravity \emph{replaces} the Einstein-Hilbert term. The resulting theory enjoys scale-invariance at high energies.  At low energies the symmetry is broken spontaneously together with the Electroweak symmetry by a suitable choice of the potential of the scalar sector \cite{Shaposhnikov:2008xi}. After symmetry breaking, the dilaton becomes a Goldstone boson and couples derivatively to any matter species in the Universe, thus avoiding the stringent constrains on the existence of a fifth-force field \cite{Kapner:2006si}. All physical scales in the low energy Lagrangian are thus given in terms of dimensionless couplings in combination with the vacuum expectation value of the dilaton. For this mechanism to lead to a vacuum state with vanishing curvature and compatible with well-known Standard Model physics, the quartic self-coupling of the dilaton must be tuned to zero. Since this coupling is related, after spontaneous breaking of scale invariance, to the value of the current cosmological constant, this creates problems with the current Dark Energy dominated era of the Universe. However, Dark Energy can be incorporated into the model by replacing standard diffeomorphism invariant gravity by \emph{Unimodular Gravity} \cite{Unruh:1988in, Henneaux:1989zc, Shaposhnikov:2008xb}, where the cosmological constant is generated as an integration constant instead of being a coupling on the Lagrangian. Apart from this subtle difference, all classical predictions of Unimodular Gravity match those of General Relativity \cite{Alvarez:2005iy,Herrero-Valea:2018ilg}.

In order to describe the power spectrum of the scalar perturbations which were produced during Higgs or Higgs-Dilaton inflation, the value of the non-minimal coupling of the Higgs field $\xi_h$ should be large, $\xi_h\sim 10^3 - 10^5 $. This fact led to an early identification of a tree-level cutoff scale which is much lower than the Planck scale \cite{Barbon:2009ya,Burgess:2010zq,Barvinsky:2009ii}. Namely, both Higgs and Higgs-Dilaton inflation seemed to enter strong coupling at energies of order $M_P/\xi_h$, close to the Hubble scale during inflation. However, in the large field domain corresponding to inflation, the cut-off scale is field dependent and it was shown in \cite{Bezrukov:2010jz} to be ultimately defined by the unitarity cut-off of gauge bosons, $\Lambda=M_P/\sqrt{\xi_h}$. This value is larger that the energy scale of inflation and cures the description of this stage within the EFT description.

Although the problem with tree-level unitarity is solved in this way, it is legitimate to feel uneasy with a coupling constant of such a large value. However, none of these feelings have found their way into a constraint or bound which is not satisfied by the model or which obstructs the large value of the non-minimal coupling. Here, we wish to follow this direction and take this study one step further, confronting HDI with the extended positivity bounds of \cite{Bellazzini:2017fep}. By rewriting the Lagrangian in a proper way, which a) gets rid of gravitational interactions by uplifting them above the Planck mass and b) shifts all non-renormalizability below $M_P$ into derivative couplings in the scalar sector; we are able to obtain a bound on the couplings of higher order operators in the EFT of the scalar fields. These bounds are easily satisfied when the non-minimal couplings take their values compatible with inflationary physics $\xi_h\sim 10^3 -10^5$ and $\xi_\sigma\lesssim 10^{-3}$.

This paper is organized as follows. First we review the derivation of posivity bounds on amplitudes in section \ref{sec:two} and introduce Higgs-Dilaton inflation in section \ref{sec:three}, where we as well rewrite the action in a suitable way for our purposes. Section \ref{sec:four} is devoted to the derivation of the bounds from the three different possible $2$-to-$2$ channels available at tree level. Finally, we draw our conclusions in section \ref{sec:five}.

\section{Amplitudes' Positivity}\label{sec:two}

In this section we review the improved positivity bounds derived in \cite{Bellazzini:2017fep}. We go over them in detail, since they will be fundamental pieces of our later work

Let us consider the 2-to-2 scattering amplitude of two scalar species $a,b \rightarrow a,b$ in the center of mass frame ${\cal M}_{ab}(s,t)$. From Lorentz invariance of the underlying theory, the amplitude will depend only on the Mandelstam variables $s$ and $t$, representing the center of mass energy and transferred momentum respectively. We will assume that we are computing this amplitude in a given EFT of a possibly unknown UV complete theory, retaining only a finite number of operators which will contribute to the scattering process, thus cutting the EFT at a given order $n$ in the cut-off expansion $O(1/\Lambda^n)$.

In the forward elastic limit $t= 0$ we can write
\begin{align}
{\cal A}_{ab}(s)={\cal M}_{ab}(s,0).
\end{align}

Note that this limit only exists for massive particles or pure contact interactions. In the case of exchange of massless particles, the $t$-channel scattering will be dominated by a pole when $t\rightarrow 0$. For scalar particles one can regulate this limit by introducing a soft mass which can be taken to vanish afterwards. For higher spin particles this mechanism does not work due to the mismatch on the number of degrees of freedom between massless and massive particles. More involved techniques are required in that case \cite{Bellazzini:2019xts,Bellazzini:2016xrt,deRham:2018qqo,deRham:2017zjm,deRham:2017avq}.

\begin{figure}
\includegraphics[scale=.4]{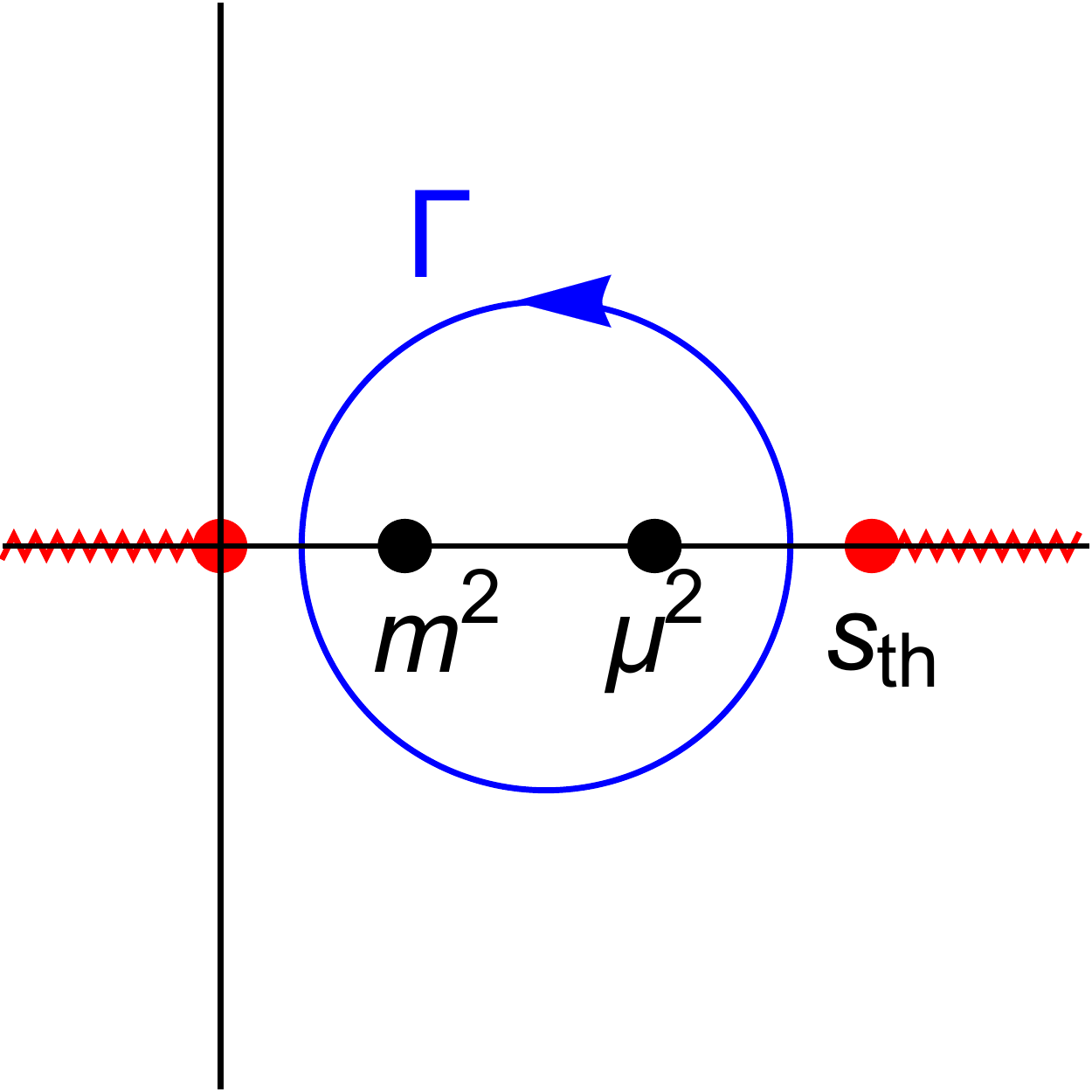}$\qquad$
\includegraphics[scale=.4]{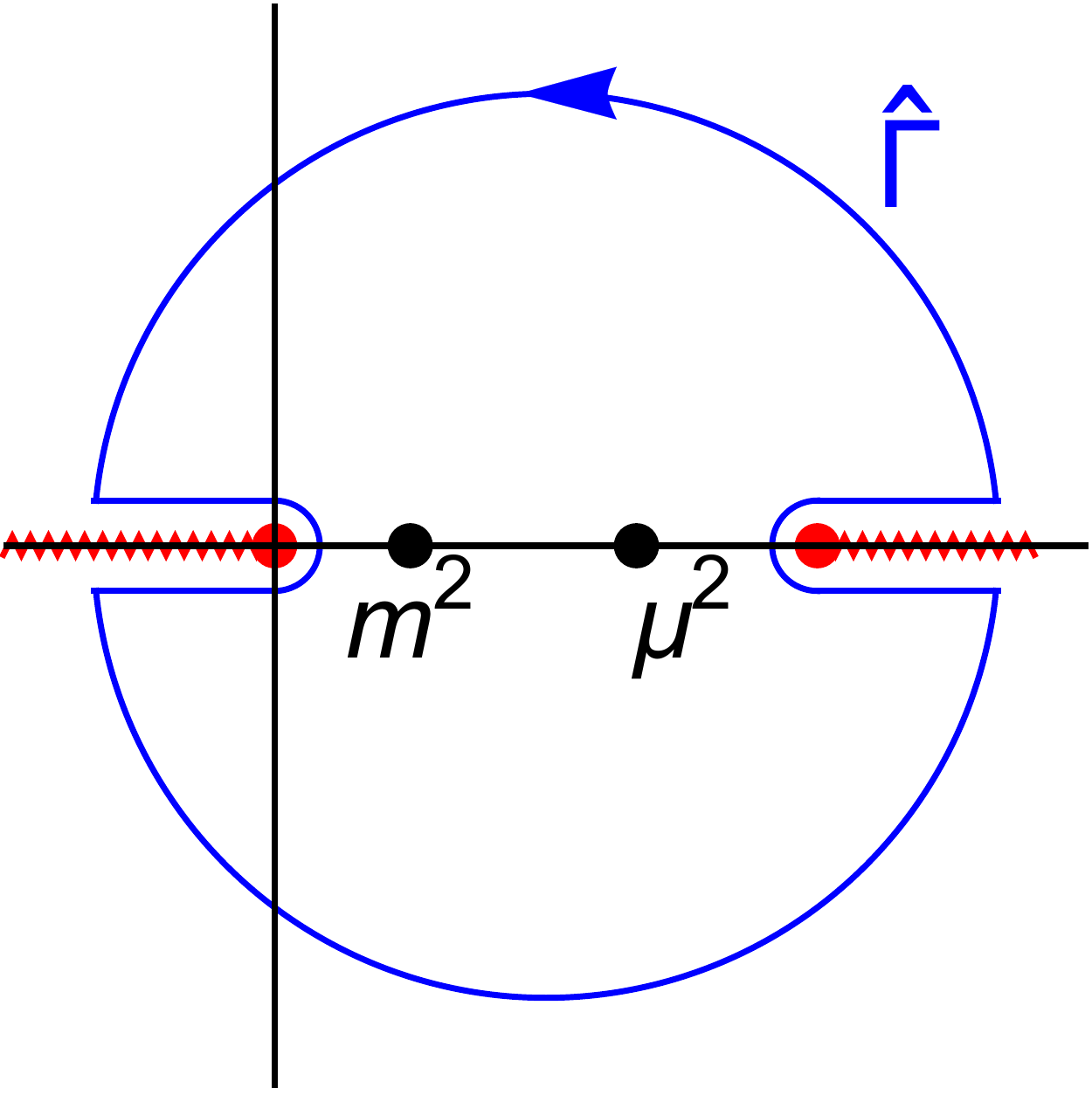} 
\caption{Integration contours in the complex s plane. We show a pole in $m^2$, being the mass of an arbitrary particle in the spectrum, as well as the pole in $\mu^2$. Left: Integration encircling all IR poles. Right: Equivalent contour in the UV theory.}\label{fig:contour}
\end{figure}

Now, following \cite{Adams:2006sv} we introduce
\begin{align}\label{sigma_pos}
\Sigma_{\rm IR}^{ab}=\frac{1}{2\pi i}\oint_{\Gamma} ds\, \frac{{\cal A}^{ab}(s)}{(s-\mu^2)^3},
\end{align}
which is evaluated in the complex plane for $s$. The contour $\Gamma$ is shown on figure \ref{fig:contour}, together with the analytic structure of the scattering amplitude. It encloses all single poles, which are determined by the masses of the particles exchanged in the $s$-channel (in the case that there are only contact interactions, these will be absent), but not the branch cut starting at $s_{\rm th}=(m_a+m_b)^2$, being $m_a$ and $m_b$ the masses of the external states, nor the one whose apex is at $s=0$ and runs over the negative real axis, produced by crossing symmetry. In the case of particle exchange, the cut starts at $s_{\rm th}=4m_i^2$ where $m_i$ is the mass of the lightest exchanged particle. The point $s=\mu^2$ is introduced to ensure convergence of the integral at $s\rightarrow \infty$ by taking into account the Froissart-Martin bound \cite{Martin:1965jj} on the scattering amplitude 
\begin{align}
\lim_{s\rightarrow\infty}|{\cal A}(s)|< {\rm constant} \cdot s (\log s)^2.
\end{align}
For convenience we can choose $\mu^2$ to lay on the real line with $\mu^2< {\rm min}\left\{(m_a+m_b)^2 , 4m_i^2\right\}$.

By the residue theorem, $\Sigma_{\rm IR}^{ab}$ is given by the sum over all the residues in the poles and thus it can be computed directly in an effective field theory (EFT) as long as all the masses are much lower than the cut-off
\begin{align}
\Sigma_{\rm IR}^{ab}=\sum_{\text{poles}} \text{Res}\left(\frac{{\cal A}^{ab}(s)}{(s-\mu^2)^3}   \right).
\end{align}

On the other hand, since there are no poles other than the ones described here, we can deform the contour to the $\hat{\Gamma}$ one shown in \ref{fig:contour}. Since the integral along the outer circle vanishes when $|s|\rightarrow \infty$, the value of $\Sigma_{\rm IR}^{ab}$ is given solely by the discontinuity across the branch cuts. Due to crossing symmetry and the fact that the amplitude is a real function of a complex variable -- thus $({\cal A}^{ab}(s))^*={\cal A}^{ab}(s^*)$ -- it can be written as
\begin{align}\label{eq:sigma_UV}
\Sigma_{\rm IR}^{ab}=\int_{s_{\rm th}^2}^\infty \frac{ds}{\pi}\left( \frac{\text{Im} {\cal A}^{ab}(s)}{(s-\mu^2)^3}  +\frac{\text{Im} {\cal A}_{\times}^{ab}(s)}{(\mu^2-u(s))^3}  \right),
\end{align}
where the second term refers to the crossed amplitude\footnote{In the case of the states $a$ or $b$ being fermionic, crossing introduces an extra minus sign. Here we deal with scalar states, so we do not need to care about the polarization basis.}
\begin{align}
{\cal A}_{\times}^{ab}(s)= {\cal A}^{ab}(u(s))
\end{align}
and $u(s)$ is the standard Mandelstam variable $u$ as a function of $s$ when $t=0$, so that $u(s)=2m_a^2 +2m_b^2 -s$. The integral is evaluated over the branch cut in the positive axis, with the lower end laying on its apex, corresponding to the threshold $s_{\rm th}^2={\rm min}\left\{(m_a+m_b)^2 , 4m_i^2\right\}$.

Equation \eqref{eq:sigma_UV} connects an infra-red quantity $\Sigma_{\rm IR}$ with the integral of the scattering amplitude \emph{up to infinite energy}, thus requiring the knowledge of the complete theory up to arbitrary high energies in order to be evaluated. It relates the behavior of the EFT with the analytic properties of the scattering amplitude derived from its UV completion. If we assume that such UV complete theory is unitary, then the imaginary part of the scattering amplitude will obey the optical theorem
\begin{align}
\text{Im} {\cal A}^{ab}(s)=s\sqrt{-u(s)/s}\,\,\sigma^{ab}(s)>0.
\end{align}

Here $\sigma^{ab}(s)$ is the total cross-section of the process, summed over all possible contributing channels
\begin{align}
\sigma^{ab}(s)=\sum_{X} \sigma^{ab\rightarrow X},
\end{align}
which is strictly positive. Thus, the same will be true for the integrand in \eqref{eq:sigma_UV} and we obtain a strict positivity bound
\begin{align}\label{eq:positivity_bound}
\Sigma_{\rm IR}^{ab}>0.
\end{align}

This will be the first result of interest that we will use later in this work. The integral $\Sigma_{\rm IR}^{ab}$ of the scattering amplitude, which can be computed within an EFT, is constrained to be positive. This will rule out already a large section of the parameter space for the couplings in front of certain operators of the EFT. But one can go further from here. One can not only argue that the integrand in \eqref{eq:sigma_UV} is strictly positive, but the different cross-sections (i.e. the different channels) that contribute to the optical theorem of a given scattering process will be positive on their own as well. This means that, in principle, we are allowed to rewrite the previous formula in the following form
\begin{align}
\Sigma_{\rm IR}^{ab}=\sum_X \int_{s_{\rm th}^2}^{\infty} \frac{ds}{\pi} \sqrt{{u(s)}{s}}\left( \frac{ \sigma^{ab\rightarrow X}(s)}{(s-\mu^2)^3}  +\frac{ \sigma_\times^{ab\rightarrow X}(s)}{(\mu^2-u(s))^3}\right),
\end{align}
where $X$ labels all possible different final states involved in the optical theorem.

In general we will not have access to all these states, since we do not know the whole spectrum of the UV complete theory. But once again, since the integrand of \emph{every term} in the sum is strictly positive, we can restrict ourselves to only those contributions given by the states present in the EFT, which give a good approximation to the scattering process of the full theory as long as our integration range is within certain energy thresholds\footnote{There is an important difference here with respect to the bounds derived in \cite{Bellazzini:2017fep}, due to the application to Higgs-Dilaton inflation described later. Since we are interested on the inflationary regime, we will consider our fields to be massless. However, this is only a good approximation as long as the energies involved in the processes satisfy $E>>m_h$ where $m_h$ is the mass of the Standard Model Higgs boson. Thus, we cannot extend the integral in \eqref{eq:beyond_positivity_bound} down to the deep infrared but instead we need to further constrain our integration regime, bounding it from below as well.} $E_{\rm IR}$ and $E_{\rm UV}$. In this case we will have
\begin{align}\label{eq:beyond_positivity_bound}
\Sigma_{\rm IR}^{ab}>\sum_{X_{\rm EFT}} \int_{E_{\rm IR}}^{E_{\rm UV}} \frac{ds}{\pi} \sqrt{{u(s)}{s}}\left( \frac{ \sigma^{ab\rightarrow X}(s)}{(s-\mu^2)^3}  +\frac{ \sigma_\times^{ab\rightarrow X}(s)}{(\mu^2-u(s))^3}\right).
\end{align}
Since all the contributions from different cross-sections are strictly positive, it does not matter how many channels we retain in the rhs of the inequality. However, the more channels, the more restricting the bound will be.

The main result of \cite{Bellazzini:2017fep} is then that $\Sigma_{\rm IR}^{ab}$ is not only constrained to be positive but to be \emph{larger than other quantity} that can be as well computed within the EFT, \emph{without knowledge of the UV completion}. This not only tells us that not all possible behaviors of the scattering amplitude are compatible with a unitary, local and Lorentz invariant UV completion. It also gives us an explicit test that we can use to further constraint our model building of EFTs. This is the form of the bound that we will exploit in the rest of this work.

\section{Higgs-Dilaton inflation}\label{sec:three}
The action for Higgs-Dilaton inflation is based on two basic assumptions: scale invariance and its spontaneous symmetry breaking. It reads
\begin{align}
S=\int d^4x \sqrt{|g|}\left(-\frac{1}{2}(2\xi_h\varphi^\dagger\varphi+\xi_\chi \chi^2)R +\frac{1}{2}\partial_\mu \chi \partial^\mu \chi -V(\varphi,\chi)\right) +S_{\rm SM}(\lambda\rightarrow 0),
\end{align}
where $\varphi$ is the Higgs boson doublet, $\chi$ is the dilaton field and $S_{\rm SM}$ is the Standard Model action with the Higgs potential removed. It is included in the potential for the combined scalar sector
\begin{align}
V(\varphi,\chi)=\lambda\left(\varphi^\dagger\varphi-\frac{\alpha}{2\lambda}\chi^2\right)^2 +\beta \chi^4.
\end{align}

Note that this action is invariant under global scale transformations of the form
\begin{align}
g_{\mu\nu}\rightarrow \Omega^2 g_{\mu\nu},\qquad \chi\rightarrow \Omega^{-1}\chi,\qquad \varphi\rightarrow \Omega^{-1}\varphi,
\end{align}
where $\Omega$ is a real constant.

For vanishing values of the dilaton self-coupling $\beta=0$, the potential leads to a vacuum state where
\begin{align}
\langle\varphi^\dagger \varphi\rangle=\frac{\alpha}{2\lambda}\langle\chi\rangle^2,
\end{align}
with vanishing space-time curvature. This also induces spontaneous breaking of both the Electroweak and scale symmetries. If $\alpha\sim 10^{-35}$ then the non-minimal coupling of the dilaton gives rise to an Einstein-Hilbert term with the right hierarchy between the Planck and Electro-weak scales, thus generating the gravitational Lagrangian at low energies. The values of the non-minimal couplings are constrained by CMB observations \cite{Bezrukov:2012hx} to be
\begin{align}
\xi_h \sim 10^3 - 10^5,\qquad \xi_\chi \lesssim 10^{-3}.
\end{align}

In the following we want to focus on the inflationary regime, thus justifying the absence of the cosmological constant in the previous action, which will dominate only at late times. We will focus on the pure inflationary sector containing $g_{\mu\nu}$, $\varphi$ and $\chi$ and we will work in the unitary gauge $\varphi=h/\sqrt{2}$ with $h$ real. The relevant action will be
\begin{align}
S=\int d^4x \sqrt{|g|}\left(-\frac{1}{2}(\xi_hh^2+\xi_\chi \chi^2)R +\frac{1}{2}\partial_\mu \chi \partial^\mu \chi+\frac{1}{2}\partial_\mu h \partial^\mu h -V(\varphi,\chi)\right).
\end{align}

In Higgs(-dilaton) models, inflation is described in the Einstein frame, given by a conformal rescaling of the metric\footnote{From now on we will restrict ourselves to tree-level scattering amplitudes, which are equivalent under field redefinitions \cite{Manohar:2018aog}. Were we considering loop corrections, a contribution from the Jacobian of the change of variables should have to be taken into account \cite{Falls:2018olk}, but this is out of the scope of this paper.}
\begin{align}
\tilde{g}_{\mu\nu}=f(h,\chi) g_{\mu\nu}, \qquad f(h,\chi)=\frac{\xi_h h^2 +\xi_\chi \chi^2}{M_P^2},
\end{align}
where $M_P$ is the Planck Mass. Using standard relations, the action becomes
\begin{align}
S=\int d^4x \sqrt{|\tilde{g}|}\left(-\frac{M_P^2}{2}\tilde{R}+\frac{1}{2}K(h,\chi)-U(h,\chi)\right),
\end{align}
with
\begin{align}\label{eq:U_potential}
U(h,\chi)=\frac{V(h,\chi)}{(f(h,\chi))^2}.
\end{align}

The kinetic term takes the form
\begin{align}
K(h,\chi)=\kappa_{AB}\tilde{g}^{\mu\nu}\partial_\mu S^A \partial_\nu S^B
\end{align}
where the kinetic metric
\begin{align}
\kappa_{AB}=\frac{1}{f(h,\chi)}\left(\delta_{AB}+\frac{3M_P^2}{2}\frac{\partial_A (f(h,\chi))^{\frac{1}{2}}\partial_B (f(h,\chi))^{\frac{1}{2}}}{f(h,\chi)}\right)
\end{align}
lives in a two-dimensional field space with coordinates $(S^1,S^2)=(h,\chi)$.

From now on, we will consider that all energies involved in our processes are within the range of validity of the EFT, i.e. they are significantly lower than the cut-off during inflation and than $M_P$ as well. Thus, for our purposes here we can decouple gravitational interactions, ending up with an EFT of two scalar degrees of freedom with non-renormalizable interactions. We will also refrain from using tildes over geometric quantities, at any time they must be interpreted as those in the Einstein frame. Moreover, we will work in the limit $\alpha\sim 0$ and thus in the approximation of massless fields, which is good as long as all energy scales involved in physical processes satisfy $E>>m_h$.
 
In this limit, the action for the scalar sector can be further simplified by performing two extra field redefinitions. First, we go to polar variables in the $(h,\chi)$ plane \cite{Bezrukov:2012hx} by
\begin{align}
\rho=\frac{M_P}{2}\log\left(\frac{(1+6\xi_h)h^2+(1+6\xi_\chi) \chi^2}{M_P^2}\right),\qquad \tan \theta=\sqrt{\frac{1+6\xi_h}{1+6\xi_\chi}}\frac{h}{\chi},
\end{align}
so that we have
\begin{align}
K=\left(\frac{1+6\xi_h}{\xi_h}\right)\frac{\partial_\mu \rho \partial^\mu \rho}{\sin^2\theta + \Z \cos^2 \theta}+\frac{(1+6 \xi_h)}{(1+6 \xi_\chi)}\frac{M_P^2 }{\xi_h}\frac{\left(\tan^2 \theta +\E\right)\partial_\mu \theta \partial^\mu \theta}{\cos^2 \theta \left(\tan^2 \theta + \Z\right)^2},
\end{align}
\begin{align}
U(\theta)=\frac{\lambda M_P^4}{4 \xi_h^2}\left(\frac{\sin^2 \theta}{\sin^2 \theta +\Z \cos^2 \theta}\right)^2.
\end{align}

The potential $U(h,\chi)=U(\theta)$ is now a function of the angular variable only. A second change of variables,
\begin{align}\label{eq:choice}
\theta=\arcsin \left(\sqrt{\frac{(1+6\xi_h)\xi_\chi \phi^2}{M_P^2 (1+6\xi_\chi)+(\xi_\chi-\xi_h)\phi^2}}\right),\qquad\rho=\varrho\,\sqrt{\frac{\xi_\chi}{1+6\xi_h}}
\end{align}
simplifies the potential to the form
\begin{align}
U(\phi)=\frac{\lambda}{4}\phi^4,
\end{align}
while the kinetic term inherits all non-renormalizability and takes the form
\begin{align}\label{eq:kinetic_final}
K=\frac{1}{2}\left(1+\frac{\xi_\chi-\xi_h}{M_P^2 
(1+6\xi_\chi)}\,\phi^2\right)\partial_\mu \varrho\partial^\mu\varrho + \frac{1}{2}\left(1+\frac{2\xi_h +6 \xi_h^2 -\xi_\chi}{M_P^2 (1+6 \xi_\chi)}\,\phi^2 + {\cal O}(\phi^4)\right)\partial_\mu\phi\partial^\mu \phi.
\end{align}

This choice of field variables is convenient for the application of the positivity bound \eqref{eq:beyond_positivity_bound} because it allows us to ignore loop contributions coming from non-renormalizable terms in the original potential \eqref{eq:U_potential}.

Note that already at this point we are finding a cut-off scale as a reflection of the non-renormalizable interactions inherited from the original gravitational Lagrangian. This can be read from the couplings of the quartic interaction in \eqref{eq:kinetic_final}, which are of the schematic form $(\rm field)^4 \partial^2/\Lambda^2$. This gives
\begin{align}\label{eq:cutoffs}
\Lambda = {\rm min}\left\{ M_P\,\sqrt{\left|\frac{1+6\xi_\chi}{2\xi_h+6 \xi_h^2 -\xi_\chi}\right|} , M_P\,\,\sqrt{\left|\frac{1+6\xi_\chi}{\xi_\chi-\xi_h}\right|}  ,\, M_P \sqrt{\left|\frac{1+6\xi_\chi}{6 \xi_h^2}\right|} \right\}.
\end{align}
where we have also added the cut-off for gauge bosons\footnote{This is the value where the tree-level scattering amplitude of two longitudinal  $W$ bosons $W_LW_L\rightarrow W_LW_L$ mediated by a Higgs boson hits the unitarity bound. } (third term) derived in \cite{Bezrukov:2012hx}, which is the most restrictive one for certain values of the non-minimal couplings. Note that in the large $\xi_h$ limit, this leads to $\Lambda\sim M_P/\xi_h$.

Let us finally remark that in \eqref{eq:kinetic_final} the coupling between $\varrho$ and $\phi$ is \emph{exact} while the self-interactions of $\phi$ come in a infinite series suppressed by higher powers of the Planck mass. Here we retain only those terms that will contribute to $2$-to-$2$ scattering, thus dropping terms with more than four fields that will not be relevant for our purposes in what follows.

\subsection{Higher derivative operators in the EFT}

\begin{figure}
\begin{fmffile}{diagrams_nonrenorm} 
\begin{fmfgraph*}(80,80) 
\fmfleft{j1,j2} 
\fmf{dashes}{j1,i}
\fmf{dashes}{j2,i}
\fmfright{f1,f2} 
\fmf{dashes,left,tension=0.4}{i,o}  
\fmf{dashes,left,tension=0.4}{o,i}  
\fmf{dashes}{o,f1}
\fmf{dashes}{o,f2}
\end{fmfgraph*} $\qquad$
\begin{fmfgraph*}(80,80) 
\fmfleft{j1,j2} 
\fmf{plain}{j1,i}
\fmf{plain}{j2,i}
\fmfright{f1,f2} 
\fmf{dashes,left,tension=0.4}{i,o}  
\fmf{dashes,left,tension=0.4}{o,i}  
\fmf{plain}{o,f1}
\fmf{plain}{o,f2}
\end{fmfgraph*} 
$\qquad$
\begin{fmfgraph*}(80,80) 
\fmfleft{j1,j2} 
\fmf{plain}{j1,i}
\fmf{plain}{j2,i}
\fmfright{f1,f2} 
\fmf{dashes,left,tension=0.4}{i,o}  
\fmf{dashes,left,tension=0.4}{o,i}  
\fmf{dashes}{o,f1}
\fmf{dashes}{o,f2}
\end{fmfgraph*} 
\end{fmffile}
\caption{Examples of one-loop logarithmically divergent diagrams generating the higher order operators shown in \eqref{eq:higher_order_ops}. Solid lines represent $\varrho$ fields, while dashed lines indicate $\phi$ fields.}\label{fig:diagrams_loop}
\end{figure}
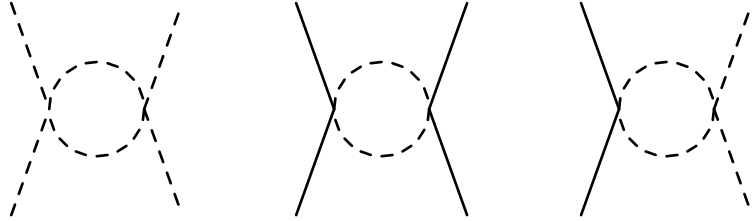

The kinetic term \eqref{eq:kinetic_final} contains non-renormalizable operators in the scalar sector. They are inherited from the non-minimal couplings to gravity in the original action. Thus, even the reduced theory in which we allow only the scalar fields to be dynamical has to be regarded as an EFT, where we expect to generate higher order operators through loop corrections. Indeed, already at one-loop there exist diagrams, depicted in figure \ref{fig:diagrams_loop}, which will generate operators with extra derivatives acting on external fields. At the leading order these will carry four derivatives and their most general form is
\begin{align}\label{eq:higher_order_ops}
\nonumber &\frac{A}{\Lambda^4}\partial_\mu \varrho \partial^\mu \varrho \partial_\nu \phi \partial^\nu \phi+\frac{B}{\Lambda^4} \partial_\mu \varrho \partial_\nu \varrho  \partial^\mu \phi \partial^\nu \phi\\
+&\frac{C}{\Lambda^4}\partial_\mu \varrho \partial^\mu \varrho \partial_\nu \varrho \partial^\nu \varrho +\frac{D}{\Lambda^4}\partial_\mu \phi \partial^\mu\phi \partial_\nu \phi \partial^\nu \phi,
\end{align}

Here $\Lambda$ will be the UV cut-off of the EFT, where the approximation given by cutting the tower of operators at a given order will break down. The Wilson coefficients $A$, $B$, $C$ and $D$ will parametrize the new physics introduced by these operators. It is worth noting that operators analogous to those in \eqref{eq:higher_order_ops} play a fundamental role in the recently proposed 
non-perturbative approach to the hierarchy problem \cite{Shaposhnikov:2018xkv,Shaposhnikov:2018jag,Shkerin:2019mmu}.

Altogether the values of these couplings are unknown from a low energy perspective. We do not know what the real cut-off of the theory is, nor the physics that we must match with the EFT. Only if we dispose of the UV complete theory, or precise experiments which measure these operators, we can say something concrete about their values. However, this setting is perfectly suitable to be confronted against the positivity bound \eqref{eq:beyond_positivity_bound}. Although we cannot know the actual value of the coefficients in the EFT, we can constrain them by requiring that they come from the low-energy expansion of a local, unitary and Lorentz invariant UV completion, as we will see in the next section. 

\section{Constraining Higgs-Dilaton inflation}\label{sec:four}
In the following we will apply the bound \eqref{eq:beyond_positivity_bound} to the higher derivative terms \eqref{eq:higher_order_ops} by looking at the three posible $2$-to-$2$ scattering processes available at tree level in HDI inflation, given by the combination of the operators in \eqref{eq:higher_order_ops} and \eqref{eq:kinetic_final}. Note that, thanks to the EFT power counting, we will not need to include loop corrections in order to apply the improved bound \eqref{eq:beyond_positivity_bound} in the case of the $\varrho \phi \rightarrow \varrho \phi$ and $\varrho \varrho \rightarrow \varrho \varrho$ channels, where all interactions will be suppressed by higher powers of either $M_P$ or $\Lambda$. Since the vertices induced by these couplings come with increasing higher powers of momenta as well, any divergent loop diagram will only renormalize higher order operators in the EFT expansion and they will not kick back into our result. This will not be the case for the $\phi \phi \rightarrow \phi \phi$ channel though, due to the presence of the renormalizable interacton $\lambda \phi^4$. In this case, loop corrections can be of the same order as the tree level contribution to the rhs of \eqref{eq:beyond_positivity_bound} and we cannot restrict ourselves to only tree-level processes. From this channel we will only be able to asses positivity of the coupling constant $D$.

\subsection{$\varrho \phi \rightarrow \varrho \phi$ scattering}
We will start by focusing on the scattering between particles corresponding to different fields, namely $\varrho \phi \rightarrow \varrho \phi$, since this channel does not receive a contribution from the potential term in the action, but it will be given solely by the kinetic term. Moreover, it will be the only channel contributing to the cross-section in the rhs of \eqref{eq:beyond_positivity_bound} as well, which makes things simpler and self-contained.

Taking together the Lagrangian \eqref{eq:kinetic_final} and the higher order operators \eqref{eq:higher_order_ops}, there is a single vertex contributing to this process
\begin{align}
\nonumber \begin{fmffile}{vertex_rprp} 
\parbox{15mm}{
\begin{fmfgraph*}(50,50) 
\fmfleft{l1,l2} 
\fmfright{r1,r2} 
\fmf{plain}{l1,c}  
\fmf{plain}{l2,c}  
\fmf{dashes}{c,r1}  
\fmf{dashes}{c,r2} 
\end{fmfgraph*}
}\end{fmffile}\quad = \frac{2(\xi_h-\xi_\chi)}{ M_P^2 (1+6\xi_\chi)}(q_1\cdot q_2)+\frac{4A}{\Lambda^4}(q_1\cdot q_2)(q_3\cdot q_4)+\frac{2B}{\Lambda^4}\left[(q_1\cdot q_4)(q_2\cdot q_3)+(q_1\cdot q_3)(q_2\cdot q_4)\right],
\label{V4}
\end{align}
where we are labelling the fields as $(\varrho_1,\varrho_2,\phi_3,\phi_4)$ and $q_i$ are their respective four-momenta. Solid lines represent $\varrho$ fields, while dashed lines indicate $\phi$ fields. All momenta are taken to be out-coming from the vertex. We are omitting factors of $i$ coming from perturbation theory.

The scattering of interest only contains $t$ and $u$ channels, with identical contributions. The scattering amplitude is then
\begin{align}
{\cal M}(s,t,u)=\frac{2A (t^2+u^2) +B (2s^2+t^2 +u^2)}{2\Lambda^4}+\frac{(t+u) (\xi_h-\xi_\chi)}{M_P^2 (1+6 \xi_\chi)}.
\end{align}

The lhs of the bound \eqref{eq:beyond_positivity_bound} can now be computed easily. Since the amplitude is at most quadratic in $s$, the value of $\Sigma_{\rm IR}$ is simply
\begin{align}
\Sigma_{\rm IR}=\frac{1}{2}\frac{\partial^2 {\cal A}(s)}{\partial s^2}=\frac{2A+3B}{2\Lambda^4}.
\end{align}

Thus, purely from positivity due to the optical theorem, we can conclude that a minimal requirement to the EFT is that
\begin{align}
2A+3B>0.
\end{align}

\begin{figure}
\includegraphics[scale=.7]{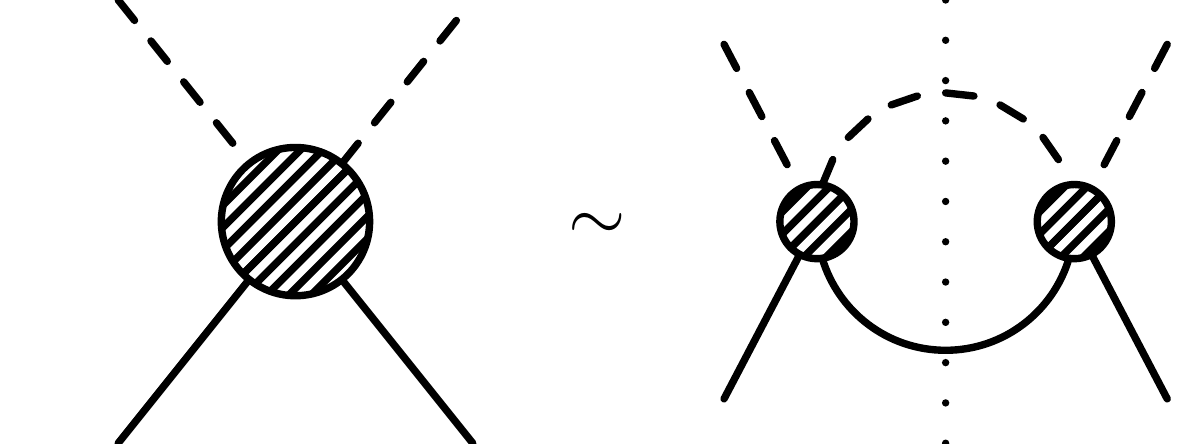} 
\caption{Schematic representation of the internal channels contributing to the cross section in the right hand side of the bound \eqref{eq:beyond_positivity_bound} for the $\varrho\phi \rightarrow \varrho\phi$ process. Solid lines correspond to $\varrho$, while dashed lines indicate $\phi$.}\label{fig:rptorp}
\end{figure}

However, the bound \eqref{eq:beyond_positivity_bound} contains even more information through the cross section of intermediate channels. Here we will consider a single channel, corresponding to the same scattering amplitude $\varrho\phi\rightarrow \varrho\phi$ as depicted in figure \ref{fig:rptorp}. The intermediate cross-section that needs to be taken into account for this process reads
\begin{align}
\nonumber &\sigma(s)=\frac{1}{16 \pi s^2}\int_{-s}^0 dt\, |{\cal M}(s,t,u(s,t))|^2 \\
&=\frac{s^3(6A^2 +11 AB +14B^2)}{480\pi \Lambda^8}+\frac{s(\xi_h-\xi_\chi)^2}{48\pi M_P^4 (1+6\xi_\chi)^2}-\frac{s^2(6A+7B)(\xi_h-\xi_\chi)}{192\pi M_P^2 \Lambda^4 (1+6\xi_\chi)}.
\end{align}

The crossed process is identical, since we are dealing with scalar particles. Thus, at leading order in $\frac{E_{\rm UV}}{\Lambda}$ and $\frac{E_{\rm IR}}{\Lambda}$, the rhs of \eqref{eq:beyond_positivity_bound} reads
\begin{align}
\int_{E_{\rm IR}^2}^{E_{\rm UV}^2} \frac{ds}{\pi} \sqrt{{u(s)}{s}}\left( \frac{\sigma(s)}{(s-\mu^2)^3}  +\frac{ \sigma(s)}{(u(s)+\mu^2)^3}\right)=\frac{(\xi_h-\xi_\chi)^2 \log\left(\frac{E_{\rm UV}}{E_{\rm IR}}\right)}{12 \pi^2 M_P^4  (1+6\xi_\chi)^2}+{\cal O}\left(\frac{E_{\rm UV}}{\Lambda},\frac{E_{\rm IR}}{\Lambda}, \frac{m_h}{E_{IR}}\right),
\end{align}
where we have restored the information that we are assuming both scalar fields to be massless in the sub-leading behavior.

Thus, we arrive at
\begin{align}
\frac{2A+3B}{\Lambda^4}\gtrsim  \frac{(\xi_h-\xi_\chi)^2 }{6 \pi^2 M_P^4  (1+6\xi_\chi)^2}\log\left(\frac{E_{\rm UV}}{E_{\rm IR}}\right),
\label{bound1}
\end{align}
which bounds the coefficient in front of the higher order interactions \eqref{eq:higher_order_ops} in relation to the cut-off $\Lambda$. This is in principle undetermined but for our purposes here we can compare it with the cut-off given by \eqref{eq:cutoffs}, which varies with the values of $\xi_h$ and $\xi_\chi$. In the particular case of Higgs-Dilaton inflation we can wonder how strict this bound is when $\xi_h>>\xi_\chi$. In that case, the lowest cut-off corresponds to the tree-level unitarity cut-off $\Lambda_{\rm tree}=M_P/\xi_h$ and we can rewrite Eq.~\eqref{bound1} as
\begin{equation}
	2A+3B \gtrsim \frac{1}{6\pi^2 \xi_h^2} \log\left(\frac{E_{\rm UV}}{E_{\rm IR}}\right).
	\label{bound2}
\end{equation} 

\begin{figure}
\begin{center}
\includegraphics[scale=.55]{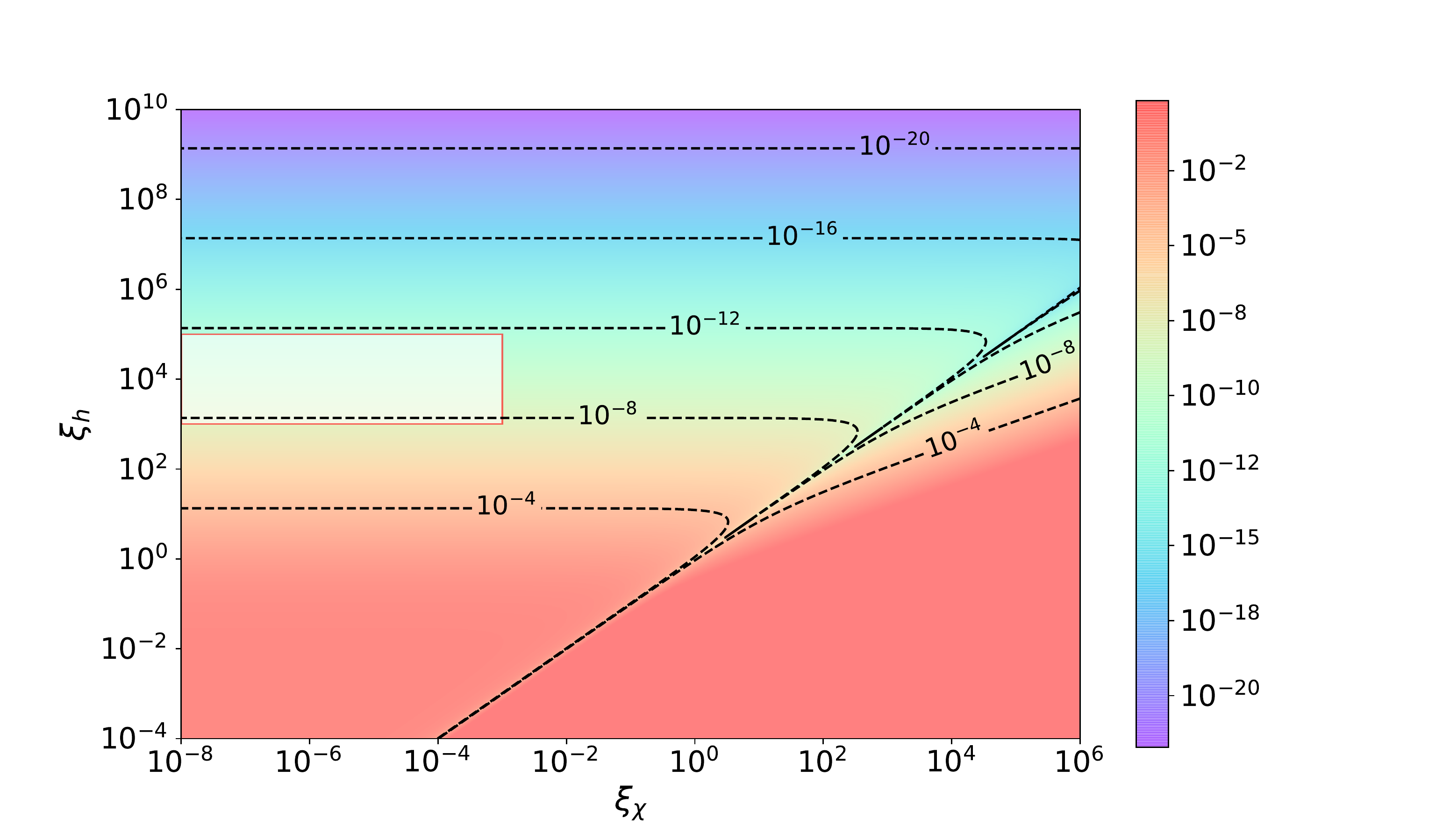} 
\end{center}
\caption{Plot of the lowest allowed value of $L_0=2A+3B$ as a function of $\xi_h$ and $\xi_\chi$. Note that both the rhs of the bound and the cut-off scale $\Lambda$ are functions of the non-minimal couplings. Level lines show different values of $L_0$. The rectangular region indicates the typical values for HDI couplings ($\xi_h \sim 10^3 -10^5$, $\xi_\chi\lesssim 10^{-3}$).}\label{fig:bound_plot}
\end{figure}

The lowest allowed value for the combination $L_0=2A+3B$ is shown in figure \ref{fig:bound_plot} for arbitrary values of $\xi_\chi$ and $\xi_h$. We observe several features. First, our bound implies that some of the operators Eq.~\eqref{eq:higher_order_ops} have to be present with non-vanishing coupling constants if the theory has a healthy UV completion. Second, the combined value $L_0$ of the Wilson coefficients is sensitive to the value of $\xi_h$ and $\xi_\chi$ through their presence both in \eqref{bound1} and \eqref{eq:cutoffs}. Although for large values of $\xi_h$, which is the relevant region for HDI, the parameter space for $L_0$ is essentially unconstrained (even though it must satisfy positivity), we find a bound of order one for small values of $\xi_h\lesssim 1 $ satisfying $\xi_h<\xi_\chi$. In the diagonal $\xi_h=\xi_\chi$ the rhs of \eqref{bound1} vanishes and we can only asses strict positivity of $L_0$.

It is worth to note that although the value of the bound is enhanced by a logarithm of the ratio between the UV and IR relevant scales, its effect is not decisive. Even in an extremal case were we could take $E_{\rm UV}=M_P$ and $E_{\rm IR}=m_h$, this leads to an enhancement of an order of magnitude only, since $\log(M_P/m_h)\sim 40$.

\subsection{$\varrho \varrho \rightarrow \varrho \varrho$ scattering}
Let us now focus on the second $2$-to-$2$ scattering of interest in our model, $\varrho \varrho \rightarrow \varrho \varrho$. In our EFT description, given by the sum of \eqref{eq:kinetic_final} and \eqref{eq:higher_order_ops}, there is a single vertex contributing to this process at tree level
\begin{align}
\begin{fmffile}{vertex_rrrr} 
\parbox{15mm}{
\begin{fmfgraph*}(50,50) 
\fmfleft{l1,l2} 
\fmfright{r1,r2} 
\fmf{plain}{l1,c}  
\fmf{plain}{l2,c}  
\fmf{plain}{c,r1}  
\fmf{plain}{c,r2} 
\end{fmfgraph*}
}\quad = \frac{8 C}{\Lambda^4}\left((q_1\cdot q_4)(q_2\cdot  q_3)+(q_1\cdot q_3)(q_2\cdot q_2)+(q_1\cdot q_2)(q_3\cdot q_4)\right)\end{fmffile}.\label{eq:V4_2}
\end{align}

The scattering amplitude, given by the combination of the $s$, $t$ and $u$ channels gives, in the forward limit
\begin{align}
{\cal A}(s)=\frac{12C s^2}{\Lambda^4},
\end{align}
which leads to
\begin{align}
\Sigma_{\rm IR}=\frac{24 C}{\Lambda^4}.
\end{align}

We thus conclude that the coefficient $C$ in \eqref{eq:higher_order_ops} must be positive in order to agree with the simple bound \eqref{eq:positivity_bound}. 

\begin{figure}
\includegraphics[scale=.7]{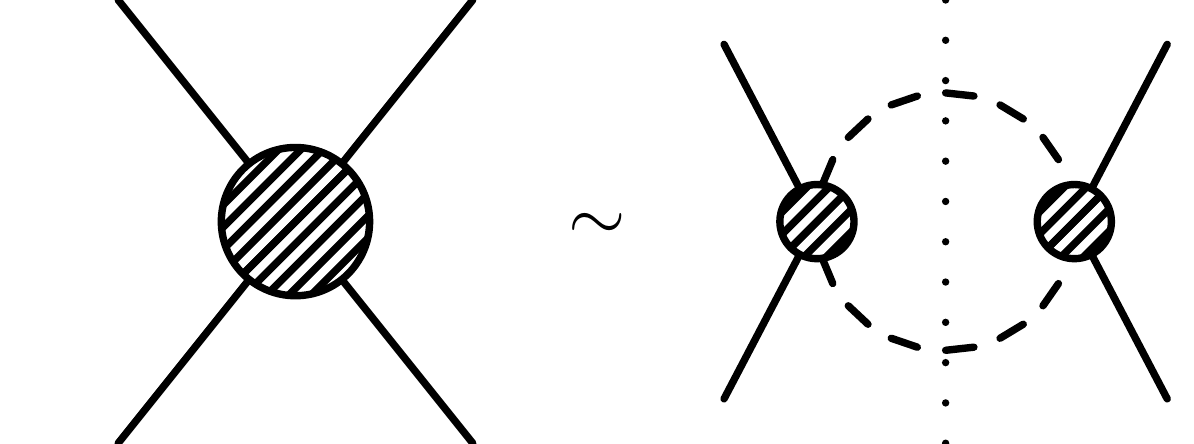} 
\caption{Schematic representation of the internal channels contributing to the cross section for the $\varrho\varrho \rightarrow \varrho\varrho$ process.}\label{fig:rrrr}
\end{figure}

In order to evaluate the improved bound \eqref{eq:beyond_positivity_bound} we need to select which channels will contribute to the rhs. In our EFT at tree level there is again a single vertex contributing to the process schematically depicted in figure \ref{fig:rrrr}, given by the s-channel of \eqref{V4}. The corresponding cross section reads
\begin{align}
\nonumber \sigma(s)&=\frac{s^3(60 A^2 +40 AB +7 B^2)}{1920 \pi \Lambda^8}+\frac{s^2(3A+B)(\xi_h-\xi_\chi)}{48 \pi M_P^2 \Lambda^4 (1+6\xi_\chi)}+\frac{s(\xi_h-\xi_\chi)^2}{32 \pi M_P^4 (1+6\xi_\chi)^2}.
\end{align}

From here we extract the rhs of the bound \eqref{eq:beyond_positivity_bound} to be
\begin{align}
\int_{E_{\rm IR}^2}^{E_{\rm UV}^2} \frac{ds}{\pi} \sqrt{{u(s)}{s}}\left( \frac{ \sigma(s)}{(s-\mu^2)^3}  +\frac{ \sigma(s)}{(u(s)+\mu^2)^3}\right)=\frac{(\xi_h-\xi_\chi)^2 \log\left(\frac{E_{\rm UV}}{E_{\rm IR}}\right)}{8\pi^2 M_P^4 (1+6\xi_\chi)^2}+{\cal O}\left(\frac{E_{\rm UV}}{\Lambda},\frac{E_{\rm IR}}{\Lambda}, \frac{m_h}{E_{IR}}\right).
\end{align}

We arrive at
\begin{align}
\frac{24 C}{\Lambda^4}\gtrsim \frac{(\xi_h-\xi_\chi)^2 \log\left(\frac{E_{\rm UV}}{E_{\rm IR}}\right)}{8\pi^2 M_P^4  (1+6\xi_\chi)^2},
\end{align}
at the leading order. Note that, since the cross section on the rhs is the same as for the previous $\varrho\phi\rightarrow\varrho\phi$, the bound is, up to a numerical factor, identical to the previous one and therefore its dependence on the value of the non-minimal couplings can be seen in figure \ref{fig:bound_plot}.

Introducing the unitarity cut-off of HDI inflation ($\Lambda\sim M_P/\xi_h$) and taking $\xi_h>>\xi_\chi$ we finally have
\begin{align}
C\gtrsim \frac{1}{192 \pi^2 \xi_h^2}\log\left(\frac{E_{\rm UV}}{E_{\rm IR}}\right).
\end{align}
\subsection{$\phi \phi \rightarrow \phi \phi$ scattering}
Finally, let us take a look at the last $2$-to-$2$ scattering available at tree level in our EFT description of the scalar sector HDI, $\phi\phi\rightarrow\phi\phi$. In this case, the interaction vertex is given by
\begin{align}
\begin{fmffile}{vertex_pppp} 
\parbox{15mm}{
\begin{fmfgraph*}(50,50) 
\fmfleft{l1,l2} 
\fmfright{r1,r2} 
\fmf{dashes}{l1,c}  
\fmf{dashes}{l2,c}  
\fmf{dashes}{c,r1}  
\fmf{dashes}{c,r2} 
\end{fmfgraph*}
}\end{fmffile}\quad =&\frac{2 \left(\xi_{\chi}^2-2 \xi_{h} (3 \xi_{h}+1)\right)
   }{M_P^2 (6 \xi_{\chi}+1)}\left((q_1\cdot q_2)+(q_1\cdot q_3)+(q_1\cdot q_4)+(q_2\cdot q_3)+(q_2\cdot q_4)
   +(q_3\cdot q_4)\right)\nonumber\\
   &+\frac{8  D }{\Lambda ^4}((q_1\cdot q_4)(q_2\cdot q_3)+(q_1\cdot q_3)(q_2\cdot q_4)+(q_1\cdot q_2)(q_3\cdot q_4))+6\lambda. \label{eq:V4_3}
\end{align}

Note a striking difference with respect to \eqref{V4} and \eqref{eq:V4_2}. While all the terms in both of these vertices were suppressed by a high-energy scale, given either by $M_P$ or $\Lambda$; in this case we have also a renormalizable interaction with no suppression at all -- the last term proportional to $\lambda$. As we will see in a moment, this term will dominate the cross section in the rhs of \eqref{eq:beyond_positivity_bound}, effectively impeding us to extract an improved bound for this scattering channel.

The amplitude for this tree level process is
\begin{align}
{\cal M}(s,t)=18\lambda+\frac{12 D \left(s^2+s t+t^2\right)}{\Lambda ^4},
\end{align}
which gives
\begin{align}\label{eq:lhs_pppp}
\Sigma_{\rm IR}=\frac{24D}{\Lambda^4}.
\end{align}

\begin{figure}
\includegraphics[scale=1]{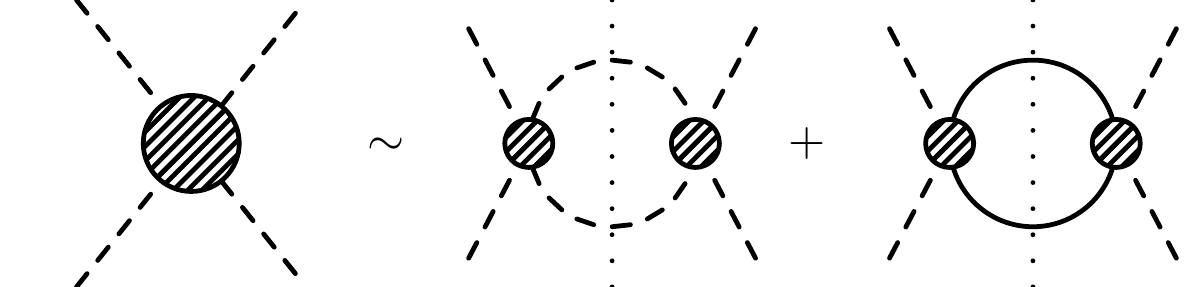} 
\caption{Internal channels contributing to the cross section for the $\phi\phi \rightarrow \phi\phi$.}\label{fig:pppp}
\end{figure}

In order to compute the rhs of \eqref{eq:beyond_positivity_bound} we can include two different channels, depicted in figure \ref{fig:pppp}. However, as previously advertised, the result will be dominated by the renormalizable part of the interaction and thus we do not need to include the second channel. Instead, we stick to the pure $\phi\phi\rightarrow \phi \phi$ interaction given by \eqref{eq:V4_3}. This gives a cross section
\begin{align}
\nonumber \sigma(s)=\frac{9\lambda^2}{8\pi s}+\frac{7 D^2 s^3}{20 \pi \Lambda^8}+\frac{5 D s \lambda}{4\pi \Lambda^4},
\end{align}
which leads to a rhs for the improved bound \eqref{eq:beyond_positivity_bound} 
\begin{align}
\int_{E_{\rm IR}^2}^{E_{\rm UV}^2} \frac{ds}{\pi} \sqrt{{u(s)}{s}}\left( \frac{ \sigma(s)}{(s-\mu^2)^3}  +\frac{ \sigma(s)}{(u(s)+\mu^2)^3}\right)=\frac{9 \lambda ^2}{8 \pi ^2
   E_{\rm IR}^4}+{\cal O}\left(\frac{E_{\rm UV}}{\Lambda},\frac{E_{\rm IR}}{\Lambda}, \frac{m_h}{E_{IR}},\frac{E_{\rm IR}}{M_P}\right).
\end{align}

Note that we are retaining the leading order in $\frac{E_{\rm IR}}{M_P}$. The rest of the terms will be suppressed by higher powers of $M_P$. This result is dominated by the renormalizable contribution of the interaction, as advertised. However, the lhs \eqref{eq:lhs_pppp} is already of order $M_P^4$ and it is suppressed with respect to this. If we wanted to give a meaningful comparison we would have to compute one-loop corrections to the lhs of \eqref{eq:lhs_pppp}, which are also dominated by the term of order $\lambda^2$ and will thus satisfy automatically the improved bound \eqref{eq:beyond_positivity_bound} thanks to the optical theorem, as shown in appendix \ref{app:1}. Therefore, for the case in which renormalizable interactions are present in the scattering channel, the improved bound \eqref{eq:beyond_positivity_bound} cannot be used at tree-level, and we can only rely on the strict positivity bound \eqref{eq:positivity_bound}
\begin{align}\label{eq:bound_49}
\frac{24D}{\Lambda^4}>0.
\end{align}

Were we studying pure Higgs inflation, this argument would have gone through in a totally identical way. In absence of the dilaton field, the interaction is dominated by the quartic self-interaction of the Higgs field, which would thus render the strong positivity bound not applicable as well. Thus, we see that the result for Higgs inflation is contained in \eqref{eq:bound_49} upon the suppression of the dynamics of $\varrho$ and proper rescaling of the coupling constants.

\section{Discussion and conclusions}\label{sec:five}
In this paper we have exploited the positivity bounds \eqref{eq:positivity_bound} and \eqref{eq:beyond_positivity_bound} on scattering amplitudes to constrain the value of the first higher derivative operators in the EFT expansion of the Higgs-Dilaton inflationary model. By demanding the unknown UV completion of the theory to be unitary, causal, local and Lorentz invariant, we have found two constraints on three of the parameters of the model, together with the positivity of a fourth one 
\begin{align}\label{eq:summary_bounds}
&\frac{2A+3B}{\Lambda^4}\gtrsim  \frac{(\xi_h-\xi_\chi)^2 }{6 \pi^2 M_P^4  (1+6\xi_\chi)^2}\log\left(\frac{E_{\rm UV}}{E_{\rm IR}}\right),\\
&\frac{32 C}{\Lambda^4}\gtrsim \frac{(\xi_h-\xi_\chi)^2 \log\left(\frac{E_{\rm UV}}{E_{\rm IR}}\right)}{6\pi^2 M_P^4  (1+6\xi_\chi)^2},\label{eq:summary_bounds2}\\
&D>0.\label{eq:summary_bounds3}
\end{align}

The scales $E_{\rm IR}$ and $E_{\rm UV}$ establish the validity of our approximations, namely that the EFT is weakly coupled within the range of energies considered, that the spectrum of states matches with the observed one and that the scalar fields can be taken as massless. Under these approximations, the logarithm in the formulas above contribute with a value ranging in $\mathcal{O}(1)-\mathcal{O}(10)$. 

The previous bounds are completely general and valid for any value of $\xi_h$ and $\xi_\chi$. However, in the particular case of HDI, the non-minimal couplings are constrained by observations to satisfy a hierarchy $\xi_h>>\xi_\chi$, while the theory has a natural cut-off given by tree-level unitarity $\Lambda = M_P/\xi_h$. If we assume this, we can simplify our bounds to be
\begin{align}
2A+3B \gtrsim \frac{1}{6\pi^2 \xi_h^2} \log\left(\frac{E_{\rm UV}}{E_{\rm IR}}\right),\qquad C\gtrsim \frac{1}{192 \pi^2 \xi_h^2}\log\left(\frac{E_{\rm UV}}{E_{\rm IR}}\right),\qquad D>0.
\end{align}
which is easily satisfied for those values of $\xi_h$ and $\xi_\chi$ compatible with CMB observations.

The obtained result is a non-trivial consistency check for the HDI model. In principle, one could expect positivity bounds to imply, say, $A,B,C>10^{4}$ which would mean that a UV completion requires large values of higher derivative operators, in tension with observations. In this case, the model enters the strong coupling regime well before the naively expected cutoff scale $M_P/\xi_h$. However, we instead obtained an extremely weak bound $A,B,C\lesssim 10^{-8}$ which means that the HDI model is a self-consistent effective field theory below the scale $M_P/\xi_h$ even if one requires that a UV completion with good properties exists.

The original form of the bounds \eqref{eq:summary_bounds}-\eqref{eq:summary_bounds3} has a larger range of validity and we expect them (or a suitable modification) to be applicable to several other physical situations. These include applications to multi-field inflation \cite{Wands:2007bd}, new Higgs inflation \cite{Germani:2010gm,Fumagalli:2017cdo}, composite models for particle physics \cite{Contino:2010rs,Brivio:2013pma}, or the recently proposed non-perturbative mechanism of generation of the separation between the Electro-weak and Planck scales \cite{Shaposhnikov:2018xkv,Shaposhnikov:2018jag,Shkerin:2019mmu}.

An important restriction of our result is the impossibility to extend the refined bounds \eqref{eq:beyond_positivity_bound} to situations with renormalizable interactions. In those cases, the rhs of the bound \eqref{eq:beyond_positivity_bound} is dominated by the contribution coming from the renormalizable couplings, while the EFT corrections are suppressed with respect to it. Accounting for this would require the inclusion of non-trivial loop corrections. This impedes us to get something better than the strict positivity bound \eqref{eq:positivity_bound} for our coupling constant $D$ and obstructs us from performing a similar study in HI, where the amplitudes are dominated by self-interactions of the Higgs boson.

Another missing feature is the study of the extended bounds under exchange of gravitons in the full theory, when gravity is dynamical. In those situations, the forward limit of the $2$-to-$2$ scattering is singular, due to a pole in the t-channel when $t\rightarrow 0$. Here we have got rid of this issue by focusing on the EFT of the scalar degrees of freedom, decoupling gravity above the range of validity of our approximations. However, it would be desirable to extend our formalism to the most general case, since it is expected that new bounds connecting non-gravitational degrees of freedom with gravitational ones will emerge, on the lines of \cite{Bellazzini:2019xts}. This would hopefully allow us to extend our results into the regime where gravitational interactions are triggered on.

\section*{Acknowledgments}
We are grateful to Sander Mooij, Francesco Riva, Valery Rubakov, Mikhail Shaposhnikov, and Andrey Shkerin for useful discussions and comments. AT thanks Alexey Koshelev for inspiring ideas on dealing with the non-renormalizable potentials. We are also very grateful to an anonymous referee for spotting an important typo in a formula. The work of MH was supported by the Swiss National Science Foundation. The work of IT was supported by the ERC-AdG-2015 grant 694896. The work of AT was supported by the Foundation for the Advancement of Theoretical Physics and Mathematics BASIS grant. The part of work of AT concerning the computation of scattering amplitudes of the Higgs and dilaton fields was supported by the  Russian Science Foundation grant 19-12-00393.

\appendix

\section{Renormalizable interactions do not go beyond positivity}\label{app:1}

As discussed in the main text, the appearance of renormalizable interactions in the Lagrangian obstructs us to go beyond positivity in certain scattering channels, due to the hierarchy between the different terms in the scattering amplitude. In order to illustrate this, let us consider a simple toy model with a single scalar field and interactions given by
\begin{align}
{\cal L}=\frac{1}{2}\partial_\m \phi \partial^\m \phi -\frac{\lambda}{4!}\phi^4 - \frac{a}{\Lambda^2}\phi^2 \partial_\m \phi \partial^\m \phi -\frac{b}{\Lambda^4}(\partial \phi)^4,
\end{align}
where we assume, in the same way as before in this text, that the scalar field has a negligible mass with respect to some infra-red cut-off $E_{\rm IR}$.

The 2-to-2 tree-level scattering amplitude given by this Lagrangian can be easily computed. In the s-channel we have
\begin{align}
{\cal A}_s=\lambda-\frac{4a}{\Lambda^2} (s-t-u) -\frac{2b}{\Lambda^4} (s^2 - t^2 - u^2),
\end{align}
and therefore the whole amplitude is
\begin{align}
{\cal A}_{\rm tree}=3\lambda +\frac{2b}{\Lambda^4}(s^2+u^2+t^2),
\end{align}
where we have used the fact that $s+t+u=0$.

If we plug this into \eqref{sigma_pos} we find
\begin{align}
\Sigma=\frac{2b}{\Lambda^4}.
\end{align}

We compute now the rhs of \eqref{eq:beyond_positivity_bound} by means of the optical theorem. The intermediate cross-section reads
\begin{align}
\sigma=\frac{2 a^2 s}{\pi \Lambda^4}-\frac{b \lambda s}{24 \pi \Lambda^4}+\frac{a b s^2}{3 \pi  \Lambda^6}-\frac{a \lambda}{2 \pi 
  \Lambda^2}+\frac{b^2 s^3}{60 \pi  \Lambda^8}+\frac{\lambda^2}{32 \pi  s}.
\end{align}

Note that most terms here are suppressed by powers of $\Lambda$ with respect to the contributions proportional to $\lambda^2$. Thus, we can write
\begin{align}
\sigma=\frac{\lambda^2}{32 \pi  s}+{\cal O}\left(\frac{1}{\Lambda^2}\right).
\end{align}

The rhs of the bound becomes
\begin{align}
\int_{E_{\rm IR}^2}^{E_{\rm UV}^2} \frac{ds}{\pi} \sqrt{{u(s)}{s}}\left( \frac{ \sigma(s)}{(s-\mu^2)^3}  +\frac{ \sigma(s)}{(u(s)+\mu^2)^3}\right)=\frac{\lambda^2}{32 \pi ^2 E_{\rm IR}^4}+{\cal O}\left(\frac{1}{\Lambda^2},\frac{1}{E_{\rm UV}^2}\right),
\end{align}
and we have, at the leading order
\begin{align}
\frac{2b}{\Lambda^4}\gtrsim\frac{\lambda^2}{32 \pi ^2 E_{\rm IR}^4}.
\end{align}

However, note that this inconsistent. In the rhs we have a contribution proportional to $\lambda^2$ while in the lhs we have used an amplitude proportional to $\lambda$. We are comparing different orders in perturbation theory. In order to write a meaningful bound we need to include contributions of order $\lambda^2$ in the lhs as well. These come from the one-loop correction to the renormalizable interaction, given by
\begin{align}
i{\cal A}_{\lambda^2}=\begin{fmffile}{one_loop_app} 
\parbox{15mm}{\begin{fmfgraph*}(60,60) 
\fmfleft{j1,j2} 
\fmf{dashes}{j1,i}
\fmf{dashes}{j2,i}
\fmfright{f1,f2} 
\fmf{dashes,left,tension=0.4}{i,o}  
\fmf{dashes,left,tension=0.4}{o,i}  
\fmf{dashes}{o,f1}
\fmf{dashes}{o,f2}
\end{fmfgraph*}}\end{fmffile}\quad + {\rm (t-channel)}+{\rm (u-channel)}.
\end{align}

In order to match with the discussion in the rest of this work, we compute this amplitude in the presence of both an infra-red and ultra-violet cut-off, that we identify with $E_{\rm IR}$ and $E_{\rm UV}$ respectively. Using standard formulas and subtracting divergences with the help of a counterterm we find
\begin{align}
i{\cal A}_{\lambda^2}^{\rm s-channel}=\frac{i\lambda^2}{32 \pi^2}\int_{0}^1 dx\,\log\left(\frac{\Delta(s) - E_{\rm UV}^2}{\Delta(s)-E_{\rm IR}^2}\right),
\end{align}
where $\Delta(n)=x(x-1)n$.

The total contribution of order $\lambda^2$ to the 2-to-2 scattering amplitude is thus
\begin{align}
i{\cal A}_{\lambda^2}=\frac{i\lambda^2}{32 \pi^2}\sum_{n=s,t,u}\int_{0}^1 dx\,\log\left(\frac{\Delta(n) - E_{\rm UV}^2}{\Delta(n)-E_{\rm IR}^2}\right).
\end{align}

Adding this to ${\cal A}_{\rm tree}$ we find that $\Sigma$ now reads
\begin{align}
\Sigma=\frac{2b}{\Lambda^4}+\frac{\lambda^2}{32\pi^2 \mu^4}+{\cal O}\left(\lambda^3,E_{\rm IR}^2,\frac{1}{E_{\rm UV}^2}\right),
\end{align}
and thus, the bound \eqref{eq:beyond_positivity_bound} reads now at the leading order
\begin{align}\label{bound_meaningless}
\frac{2b}{\Lambda^4}+\frac{\lambda^2}{32\pi^2 \mu^4}\gtrsim \frac{\lambda^2}{32 \pi ^2 E_{\rm IR}^4}.
\end{align}

However, our starting assumptions include $\mu <4m^2$ and $E_{\rm IR}>4m^2$ in order to be able to neglect the mass of the field. We thus have
\begin{align}
E_{\rm IR}>\mu,
\end{align}
and the bound \eqref{bound_meaningless} \emph{is always satisfied regardless of the value of $b$ and $\Lambda$}. Therefore, we cannot go beyond positivity or, better said, going beyond positivity does not give any new information. We can thus only asses pure positivity.

Note that this is a consequence of the presence of renormalizable interactions. Due to the optical theorem, the rhs of the bound must be always smaller than the lhs. Only in the case in which we could extend the EFT up to arbitrary energies and down to $E_{\rm IR}=2m$, we would saturate the equality, as corresponds to a purely renormalizable theory.

In the case of non-renormalizable interactions, the vertices and the scattering amplitudes are suppressed by a cut-off scale $\Lambda$ and therefore the vertex squared is \emph{not} the leading contribution to the rhs of the bound anymore and one does not need to compute loop contributions. This can be also understood from the standard knowledge that in an EFT, loop corrections renormalize higher order operators only.

\bibliography{positivity}{}

\end{document}